\documentstyle[aps,preprint]{revtex}
\begin{document}
\draft
\tighten
\preprint{\vbox{Submitted to Physical Review C
                \hfill YUM 96-23\\
\null \hfill  SNUTP 96-106}}
\title{Delta Decay in the Nuclear Medium}
\author{Hungchong Kim $^1$\footnote{Email : hung@phya.yonsei.ac.kr}, 
S. Schramm $^{2}$\footnote{Email : schramm@tpri6e.gsi.de}, 
Su Houng Lee $^1$\footnote{Email : suhoung@phya.yonsei.ac.kr}}
\address{$^1$ Department of Physics, Yonsei University, Seoul, 120-749,
Korea \\
$^2$ GSI, D-64220 Darmstadt, Germany} 

\maketitle
\begin{abstract}
The $\Delta$ decay in the nuclear medium is calculated in the relativistic
meson-nucleon model. The delta spreading width is calculated and compared
with the Pauli-blocked $\pi$N decay width.  
The influence of relativistic mean fields is also studied.
We stress the importance of understanding the delta spreading
width in interpreting experiments involving delta resonances.
\end{abstract}  

\section{INTRODUCTION}
\label{sec:intro}
The understanding of the properties of the $\Delta$ resonance in 
the nuclear medium, especially the branching ratios of its various decay 
modes, is important for the analysis of scattering processes
involving $\Delta$-hole excitations.  

A $\Delta$, which mainly decays into 
$\pi$N in free space, acquires additional decay channels in the medium. 
For example, it can decay through the
mechanism, $\Delta + {\rm N} \rightarrow {\rm N} + {\rm N}$, 
commonly known as
``delta spreading width'' which can be seen as reabsorption of 
the virtual pion through a particle-hole excitation of the nucleus.
Indeed, a number of experiments involving $(p, n)$ reactions~\cite{had}  
show that the additional channels of delta decay seem to dominate over 
the usual $\pi$N decay.  

Recently, we have proposed that the spreading width could be
the source of errors in measuring the nucleon axial form factor from 
charged-current neutrino-nucleus scattering~\cite{delta}.  Namely, 
in the BNL experiment of neutrino-nucleus scattering~\cite{BNL} where 
neutrino beams with an average energy of 1.3 GeV are used, the data are 
usually fitted to the theoretical curve calculated in a Fermi gas model of 
the nucleus in order to extract the axial form factor.  In this approach  
$\Delta$-hole excitations are assumed to be completely excluded by   
experimentally
rejecting events involving pions in the final state.   
However, when the $\Delta$ obtains a non-zero spreading width,  some events 
which generate $\Delta$-hole excitations lead to  pionless decays 
of the $\Delta$ which cannot be excluded by merely detecting the pions.
To estimate these events quantitatively,  it is important to know the 
branching ratio of the spreading width to the $\pi$N decay width 
before one can deduce the nucleon axial form factor accurately.

Further, the knowledge of the branching ratio can be used to improve 
Monte Carlo simulations of neutrino detectors, 
e.g. in the Kamiokande experiment, where the flavor ratio of atmospheric
neutrinos are measured~\cite{kamiokande}.  It was suggested that pions 
produced from the charged-current neutrino-nucleus reactions 
cause  uncertainties in identifying the flavor of the neutrinos 
created in the earth's atmosphere~\cite{atmos}.   
In these experiments the main source of pions is believed to
come from delta decays.  Therefore, a thorough knowledge of
the branching ratio 
between pionic and non-pionic decays of deltas is crucial for estimating
the pion events in neutrino-nucleus scattering of atmospheric
neutrinos.

In this work, we present a  calculation of the delta 
decay width and its modification in nuclear matter 
within the framework of the Walecka model~\cite{brian}. The model   
is a relativistic meson-nucleon model which has been quite successful 
in describing nuclear properties, including 
response functions~\cite{horowitz}.   Crucial ingredients of this model are 
the presence of large vector and scalar self-energies which modify the 
nucleon properties in the medium enhancing relativistic effects.  Our 
concern is to study the modification of the delta decay width driven by these
relativistic effects.  Nonrelativistic calculations of the delta decay in the
medium can be found in Refs.~\cite{oset,arve}.

\section{Formalism}
\label{sec:pin}

In this section, we present the relativistic calculation of the 
$\Delta$ decay in the nuclear medium.  First we review the calculation of 
the $\pi$N decay which has been studied in Ref.~\cite{pauli}. Then 
we will present our calculation of the spreading width and its modification 
due to nuclear environment.

In free space, a delta decays to $\pi$N.   The decay width for this mode 
can be obtained from the imaginary part of the delta self-energy as shown 
in Fig.~\ref{self}~(a).  Formally, the decay width is given in terms of
the self-energy $\Sigma_{\mu\nu}$ and the Rarita-Schwinger spinor of the 
delta, $\Delta^\mu$,

\begin{equation}
\Gamma_f = -2\ {\rm Im} ({\bar \Delta}^\mu \Sigma_{\mu\nu} \Delta^\nu)\ 
\label{freeself}
\end{equation}
To calculate the self-energy, we introduce the $\pi {\rm N} \Delta$
interaction Lagrangian,
\begin{equation}
{\cal L}_{\pi {\rm N} \Delta} = {f_\Delta \over m_\pi}
{\bar \Delta}^\mu\  {\bf T}\ {\rm N}\ 
\partial_\mu {\mbox{\boldmath $\pi$}} + {\rm h. c.} \ ~~ .
\label{del_lag}
\end{equation}
Here  N, {\mbox{\boldmath $\pi$}}, and $\Delta$ are the nucleon, pion and
delta field, respectively.
The $2\times 4$ isospin matrices ${\bf T}$ satisfy~\cite{tjon}
\begin{equation}  
T^i (T^\dagger)^j =\delta^{ij}-{1\over 3} \tau^i \tau^j\ .
\end{equation}             

There is some ambiguity in the form of the Lagrangian of a spin-3/2 
field, which affects the off-shell behavior of the particle.
We will not discuss this complication in the following because
the delta is on-shell in our calculation.
Using this Lagrangian and standard Feynman rules, one writes the self-energy 
in free space,
\begin{equation}
\Sigma_{\mu\nu} (q) = \left( {f_\Delta \over m_\pi}
\right)^2 i \int {d^4 k \over (2\pi)^4} q_\mu q_\nu D_0 (q) G_f(k)\ ,
\label{self_f}
\end{equation}
where the pion [$D_0 (q)$] and nucleon [$G_f(k)$] propagators in free space
are given as
\begin{eqnarray}
D_0 (q) &=& { 1 \over q^2 - m_\pi^2 + i \epsilon} \label{fpion}\ ,\\
G_f (k) &=& { \not\!k + m \over k^2 - m^2 + i \epsilon}\ .
\end{eqnarray}
Putting these into Eq.~(\ref{freeself}), one can easily calculate the
free decay width,
\begin{eqnarray}
\Gamma_f &=&  \left({f_\Delta \over m_\pi} \right)^2 
{(E_{\bf k}+m)\over 12 \pi m_\Delta} |{\bf k}|^3\label{free}\ ,\\
{\rm with}&& \nonumber \\
{\bf k}^2 &=& \left({m_\Delta^2 + m_\pi^2 -m^2 \over 2 m_\Delta}\right)^2 
-m^2_\pi
\;; \quad E_{\bf k} = \sqrt{{\bf k}^2+m^2}\ .\nonumber
\end{eqnarray}
Here $m$, $m_{\Delta}$, and $m_\pi$ represent the mass of nucleon, delta and
pion respectively.  By comparing Eq.~(\ref{free}) with
its experimental value of $115$ MeV, one can fix  the coupling constant 
$f_\Delta = 2.15$. 
  
In the nuclear medium viewed as a free Fermi gas of nucleons, 
the $\pi$N decay is suppressed by  Pauli blocking, since 
the phase space available to the decaying nucleon
is restricted by the occupied Fermi sea. 
Indeed, at nuclear saturation density the $\pi$N decay is completely
blocked for a delta at rest in the medium.  When the delta
is moving, however, the decay sphere of the nucleon in momentum space becomes
an ellipsoid, which only 
partially overlaps with the Fermi sphere~\cite{pauli}. 
Therefore, the delta obtains a non-zero energy-dependent width. 

The Pauli blocking can be incorporated by replacing  
$G_f (k)$ with 
\begin{equation}
G_0(k) = G_f(k) + 2 \pi \delta(k^2 -m^2) \Theta(k_0) \Theta (k_{\rm F} 
- |{\bf k}|) (\not\!k + m)\ .
\end{equation}
Here $k_{\rm F}$ is the Fermi momentum whose value at the nuclear saturation 
is 1.3 fm$^{-1}$.  Note that in this form the nucleon propagator
is given in the rest frame of the nuclear matter where
the delta is moving with a certain relative velocity.
From the following identity~\cite{tjon} for the $\Delta$ spinor with
momentum $p^\mu=(E_\Delta, {\bf p}_\Delta)$
\begin{eqnarray}
\sum_\sigma \Delta^\mu (\sigma) {\bar \Delta}^\nu (\sigma) &=& 
-{\not\!p_\Delta + m_\Delta \over 2 m_\Delta } \left[ g^{\mu\nu} - {2 \over 3} 
{p^\mu_\Delta p^\nu_\Delta \over m^2_\Delta} + {1 \over 3 m_\Delta} 
(p^\mu_\Delta \gamma^\nu - p^\nu_\Delta \gamma^\mu) - 
{1 \over 3} \gamma^\mu\gamma^\nu \right]\nonumber \\
&\equiv& -{\not\!p_\Delta + m_\Delta \over 2 m_\Delta } 
{\rm P}^{\mu\nu}_{3/2}\ ,
\end{eqnarray}
the decay width averaged over the delta spin becomes
\begin{eqnarray}
{\bar \Gamma}_{\pi {\rm N}} (|{\bf p}_\Delta|) =  - {1 \over 8 m_\Delta
} \left({f_\Delta \over m_\pi}\right)^2 \int {d^4 k \over (2\pi)^2}
&&\Theta (q_0)\delta(q^2-m_\pi^2)\Theta(k^0- E_F) \delta(k^2-m^2)
\nonumber \\
\times && {\rm Tr} [ q_\mu q_\nu (\not\!k + m) (\not\!p_\Delta + m_\Delta)
{\rm P}^{\mu\nu}_{3/2}]\ .
\end{eqnarray}
${\bf p}_\Delta$ is the delta momentum relative to the 
nuclear matter, corresponding to a relative 
velocity $v = {|{\bf p}_\Delta|\over E_\Delta}$. Note that we have made use of
Cutkosky rule~\cite{cut} in deriving this.  
The integration over the nucleon momentum $k$ leads to 
\begin{eqnarray}
{\bar \Gamma}_{\pi {\rm N}} (|{\bf p}_\Delta|) &=& {1 \over 192 \pi}
\left({f_\Delta \over m_\pi}\right)^2
{E_2 \over |{\bf p}_\Delta| m^3_\Delta}
[(m_\Delta + m_\pi)^2 - m^2]
\nonumber \\
&\times& [(m_\Delta-m_\pi)^2-m]
[(m_\Delta + m)^2 - m^2_\pi]\nonumber\ ,\\
{\rm where} &&\ \\
E_2 &=&\Bigl[{\rm min} (E_{\Delta}, E_+)-{\rm max} (E_F, E_-)\Bigr ]\
\Theta (E_2) \nonumber\ ,\\
E_{\pm} = &{\rm max} \Biggr \{ & E_{F},{
(m^2_\Delta+m^2-m^2_\pi)  E_\Delta \pm
|{\bf p}_\Delta|\sqrt{[(m_\Delta-m)^2-m^2_\pi][(m_\Delta + m)^2 - m^2_\pi]}
\over 2 m^2_\Delta}\Biggl \}\nonumber\ .
\end{eqnarray}
A similar calculation has also been done in Ref.~\cite{pauli}.  
The authors of Ref.~\cite{pauli} also discussed 
the results for different magnetic quantum numbers of the delta, which is
qualitatively similar to the averaged result.

The free Fermi gas calculation can be improved by incorporating
mesonic mean fields.  In a mean-field approximation to the
Walecka model,  the propagation of the delta and the nucleon is modified
by the presence of constant scalar and vector mean fields which are 
determined from nuclear saturation properties.   

The $\pi$N decay width in this approximation can be obtained with 
the replacements
\begin{eqnarray}
m_\Delta \rightarrow m^*_\Delta &=& m_\Delta - S_\Delta\nonumber\ ,\\
E_{\Delta} \rightarrow E_{\Delta}^* &=& \sqrt{m^*_\Delta+{\bf p}^2_\Delta}
\nonumber\ ,\\
m \rightarrow m^* &=& m-S_{\rm N}\nonumber\ ,\\
k^\mu \rightarrow k^{*\mu} &=&k^\mu - V^\mu_{\rm N}\nonumber\ ,\\
p^\mu_\Delta \rightarrow p^{*\mu}_\Delta &=& p^\mu_\Delta - V^\mu_\Delta
\label{mean}\ .
\end{eqnarray}
The nucleon self-energies ($V^\mu_{\rm N}$, $S_{\rm N}$) are assumed to
be the same as the
delta self-energies ($V^\mu_\Delta$, $S_\Delta$) as is motivated from
the universal coupling assumption.  In symmetric nuclei,
the space components of the vector self-energy
are averaged to zero.   Also its time component can be eliminated  
by appropriately shifting integration variables.  However, the scalar
self-energy $S_{\rm N}$ does contribute.  Its value at a given nuclear 
density can be determined
by solving the self-consistency equation~\cite{brian}
\begin{equation}
m^*= m - {g_s^2 \over m_{s}^{2}} {4 \over (2\pi)^3}
 \int^{k_F}_{0} d^3 k {m^* \over
\sqrt{{\bf k}^2+{m^*}^2}}\label{self_con}\ ,
\end{equation}
where the ratio of the scalar meson coupling to its mass, $g_s^2 / m_{s}^{2}$,
is determined from the equilibrium properties of nuclear matter~\cite{brian}.
At nuclear 
saturation of $k_{\rm F}=1.3$ fm$^{-1}$, Eq.~(\ref{self_con}) yields 
$m^*=510$ MeV.

In addition to the $\pi$N decay, a delta in the medium can also
decay into 2p-1h through
the process, $\Delta + {\rm N} \rightarrow {\rm N} + {\rm N}$. The lowest-order self-energy diagram for this process is shown in Fig.~\ref{self}~(b).
This spreading width can be calculated in the same way as the $\pi$N decay 
by replacing the free pion propagator with the pion 
propagator $D(q)$ including particle-hole polarization:
\begin{equation}
D (q) = { 1 \over q^2 - m_\pi^2 - \Pi(q) + i \epsilon} \label{cpion}\ .
\end{equation}
We proceed by introducing the interaction Lagrangian for nucleons and pion
using the pseudovector coupling for the pion, 
\begin{equation}
{\cal L}_{\pi {\rm NN}} = {f_{\rm N}\over m_\pi}
{\bar {\rm N}} \gamma^\mu \gamma_5 
{\mbox{\boldmath $\tau$}} {\rm N}\ \partial_\mu {\mbox{\boldmath $\pi$}}\ ,
\label{n_lag}
\end{equation}
with $f_N = 1.01$ .
Using the interaction Lagrangian Eq.~(\ref{n_lag})
we can easily determine the lowest order particle-hole polarization
$\Pi_{\rm ph}(q)$:
\begin{eqnarray}
{\rm Re}\, \Pi_{\rm ph}(q) &=& \left(\frac{f_N}{m_\pi}\right)^2
{q^2 m^2 \over \pi^2 |{\bf q}|}
\int_m^{E_{F}}  d E_{\bf k} \log\left| 
{ (q^2 - 2 |{\bf k}| |{\bf q}|)^2 - 4 q_0^2 E_{\bf k}^2 
\over (q^2 + 2 |{\bf k}| |{\bf q}|)^2  - 4 q_0^2 E_{\bf k}^2 } 
\right| \\
{\rm Im}\, \Pi_{\rm ph}(q) &=&  \left(\frac{f_N}{m_\pi}\right)^2 q^2
{m^2 E_1\over  \pi |{\bf q}|} 
\end{eqnarray}
with
\begin{eqnarray}
E_1 = E_{F} - E_- ~~;~~ E_- = {\rm min} (E_{F},E_{\rm max}) \\
E_{\rm max} = {\rm max} \left[ m,\ E_{F}-q_0,\ -{q_0 \over 2} +
{|{\bf q}| \over 2} \sqrt{1-{4 m^2 \over q^2}} \right]\ .
\end{eqnarray}
The $\Delta$ decay calculated with Eq.~(\ref{cpion}) contains two
decay channels. First, the contribution from the usual 
$\pi N$ decay which can be identified with the pole contribution 
of the pion propagator Eq.~(\ref{fpion}) with the shifted mass
\begin{equation}
\tilde{m}_\pi^2 = m_\pi^2 + \Pi(q^2=\tilde{m}_\pi^2)\label{shift}\ . 
\end{equation}
If the lowest-order 
nucleon-hole polarization is used for $\Pi (q)$, then, in some 
kinematics, Eq.~(\ref{shift}) is solved by 
negative $\tilde{m}_\pi^2$ indicating a tachyonic pole.  This 
pole can be removed by introducing the Landau-Migdal parameter $g'$
which is incorporated by writing~\cite{hert}
\begin{eqnarray}
\Pi(q) = {q^2 \Pi_{\rm ph} \over q^2 + g' \Pi_{\rm ph}}\label{c-pol}\ .
\end{eqnarray}
In our calculation,  we take a conventional value of 
the Landau-Migdal parameter, $g' = 0.7$.  
One has to be a little bit careful with the pole contribution
since the residue of the propagator is now changed due to 
the real part of $\Pi (q)$. 
However, the modification of $\pi$N decay due to the real part of 
Eq.~(\ref{c-pol}) is very small.

The second contribution arises from the imaginary part
of the polarization Eq.~(\ref{c-pol}). This part can be identified
with the particle-hole decay channel of the virtual pion
generating the spreading width of the $\Delta$.
Note that the imaginary part of the particle-hole polarization
is only non-zero at space-like momenta of the pion.  Therefore, 
the 
spreading width can be  separated from the $\pi$N decay in the medium due
to their different kinematic region.
Replacing $D_0(q)$ by the expression Eq.~(\ref{cpion})  in Eq.~(\ref{self_f})
and taking the contribution arising from the 
imaginary part of the polarization,  
we
obtain the spreading width in the nuclear rest frame,
\begin{eqnarray}
{\bar \Gamma}_s (|{\bf p}_\Delta|) &=& {1 \over 12 m^3_\Delta \pi^2}
                   \left({f_{\Delta} \over m_\pi} \right)^2
                       \int^{E_\Delta}_{E_F} d E_k
                       \int^1_{-1} d {\rm cos}\theta\nonumber \\
                      &&\times { {\rm Im} \Pi(q) \Theta(-q^2)
                        \over [q^2-m^2_\pi -{\rm Re}\Pi(q)]^2
                        + [{\rm Im}\Pi(q)]^2}\
                        (k\cdot p_\Delta + m\ m_\Delta)
                        \Bigl [q^2 m^2_\Delta - (p_\Delta \cdot q)^2
\Bigr ]\label{sdecay}\ , 
\end{eqnarray}
which can be solved numerically. The lowest order
spreading width shown in Fig.~(\ref{self})~(b), which is usually referred to
as the spreading width, can be obtained from Eq.~(\ref{sdecay})
by taking the lowest-order contribution from the polarization, only,
and by neglecting the Landau-Migdal correction. 
Note that, unlike in the case of the free $\pi$N decay, 
the pion momentum, $q^2$, is not restricted
to its on-shell value in the course of the integration. Therefore, 
one should consider the momentum dependence of 
the $\pi$N$\Delta$ and $\pi$NN vertices.  
In this regard, we employ the monopole form for the form factor,
\begin{equation}
F(q^2) = {\Lambda^2 - m_\pi^2 \over \Lambda^2 - q^2}\ , 
\end{equation}
in our discussion.  This is to take into account the finite size of the 
nucleons as well as the short-range effects of NN interactions.  However, 
the value of the cut-off parameter $\Lambda$ is in doubt.  In principle, this
should be determined by summing the vertex diagrams 
of the underlying field 
theory.  Here we use the Bonn potential values~\cite{bonn}; for
$\pi$NN, $\Lambda = 1.8$ GeV and for $\pi$N$\Delta$, $\Lambda = 0.85$ GeV.
Finally, in order to implement the effects due
to the nuclear self-energies, the replacements in Eq.~(\ref{mean}) are 
done for the calculation of the spreading width. 

In addition to the particle-hole polarization,
$\Delta$-hole excitations can contribute. This means
that a delta rescatters in the medium by $\Delta + N \rightarrow \Delta + N$,
which can be obtained by taking the imaginary part of $\Delta$-h polarization.
The difference to the particle-hole polarization is that the imaginary part
has finite values even for positive $q^2$.  Depending on the four-momentum 
the pion pole is shifted into the complex
plane as the pion obtains a width from the decay channel to a
$\Delta$-h excitation in the medium.  The rescattered $\Delta$  
subsequently decays leading to  more complex decay
channels, for example 2N - 1h - 1$\pi$, etc.
The possible influence of this contribution will be studied in a forthcoming
publication~\cite{future}.

\section{Result}
\label{sec:result}
We now present the result for the various delta widths.  
Our concern is to understand the momentum dependence of the decay width as
well as the role of the nuclear mean fields.   

In the following, we will show our results calculated at
half, normal and twice the  nuclear saturation density with the corresponding
Fermi momenta  $k_F=203$ MeV, $256$ MeV, and $323$ MeV, respectively.
The self-consistency equation of Eq.~(\ref{self_con}) yields an 
effective nucleon mass $m^*=510$ MeV  at
the saturation density, $m^* = 715$ MeV at half saturation density,
and $m^* = 261$ MeV at twice nuclear density.  
The reduction of the nucleon mass in the medium  arises due to the
strong scalar self-energy.  In principle, 
the self-energies are functions of 
the momentum of the propagating nucleon~\cite{momen}.  Specifically,
the absolute values of the self-energies decrease 
as the momentum increases.  This momentum dependence
could be interesting especially  in the high momentum region. However, 
in most of the momentum range of our
concern, the momentum dependence of the self-energies is not expected to
be important. The result with half the
saturation density is applicable to ($p$, $n$) reactions where
the interactions most likely occur on the nuclear surface.
On the other hand, the result at saturation density could be 
used  in lepton-nucleus scattering as the incoming lepton can probe
the region deeply inside the nucleus.   The result with higher density could be 
interesting in the heavy-ion collision even though the neglect of
contributions from the lower continuum of the nucleons
in our calculation is questionable. 

Figures~\ref{wid} show the results for the $\pi$N decay width as a function
of the delta momentum $|{\bf p}_\Delta|$. 
Fig.~\ref{wid}~(a) shows the result of free Fermi gas calculation while
Fig.~\ref{wid}~(b) includes the nuclear mean fields.  
At  saturation density or at twice the saturation density,
the decay width  increases as the delta momentum increases.
This is because the overlap between the momentum ellipsoid of the 
decay nucleon and the Fermi sphere is getting smaller and the
Pauli blocking becomes less important. 
Indeed, as is shown by the dashed curve, 
the ellipsoid is completely out of the Fermi sphere for
$|{\bf p}_\Delta| > 650$ MeV, and the $\pi$N decay width is the same as
its value in free space. 
A similar trend can be observed for the mean field calculation
in Fig.~\ref{wid}~(b).  
At half saturation density (the solid curve in both figures), 
however, Pauli blocking is less significant as could be expected.  
Our results indicate that the mean field combined with
the Pauli blocking reduces the $\pi$N decay width 
substantially from its free value of 115 MeV 
which agrees with Wehrberger and Wittman~\cite{pauli}.

Unlike the $\pi$N decay, the spreading width, which is represented by the
mechanism
$ \Delta + {\rm N} \rightarrow {\rm N} + {\rm N} $, is plotted
in Figs.~\ref{swid}.  Figure~\ref{swid}~(a) shows our calculated spreading
width without the mean fields and Figure~\ref{swid}~(b) 
includes the mean fields.  First note that the momentum dependence of the 
spreading width is not as significant as the $\pi$N decay. 
This means, when added to the $\pi$N decay width,
the momentum dependence of the in-medium decay width 
is mainly driven by the Pauli blocking.

Conventionally the spreading width is scaled as the
nuclear density~\cite{moniz}.  This is reflected in the Fermi
gas results in Fig.~\ref{swid}~(a) where the dashed curve is about
a factor of 2 smaller than the dot-dashed curve and the
solid curve is about a half of the dashed curve.  
But this kind of scaling is not applicable in the mean field case.
Therefore, scaling 
the spreading width with density has to be applied with caution. 
Our mean field results at the saturation density  which vary
from 75 MeV to 57 MeV qualitatively agrees with the 
conventionally adopted value of 70 MeV in the nonrelativistic 
calculations~\cite{moniz,jain}. Also at this density, the branching ratio
of pionic decay to the spreading width, $\Gamma_{\pi {\rm N}}/\Gamma_s$,
varies from 0 to 1.33 indicating that a significant portion of deltas
leads to the pionless decays.  However the role of the spreading width 
compared to the $\pi$N decay becomes smaller at  half saturation density. 

On the other hand, the mean field result for the spreading width
at the double saturation density is rather abrupt.   Its magnitude is
only 6 $\sim$ 12 \% of the Fermi gas result. This kind of huge reduction
due to the mean fields 
might come from neglecting the vacuum contribution to the particle-hole
loop in Hartree approximation.  At this density in our simple calculation, the
effective mass of nucleon is 261 MeV, much smaller than its free value. 
Therefore, anti-nucleons can be easily excited which could lead to 
the appreciable contribution from the vacuum.  More work on the vacuum
contribution should be done~\cite{future}.  
However, as indicated in Fig.~\ref{swid}~(a) and (b), the spreading width
has a very strong dependence on the mean field given our approximations.
 
$\Delta$-hole contributions in neutrino-nucleus scatterings have been discussed
in Ref.~\cite{delta,atmos}. There it was stressed that the pionic
versus non-pionic decay could be an important source of uncertainty 
in determining the nucleon axial form factor or estimating the pion events
in atmospheric neutrinos.  For example, in Ref.~\cite{delta}, we have found 
that the theoretical error of about 0.1 GeV in the extracted axial mass 
$M_A$ is expected from the events due to delta-hole excitations. 
This conclusion is
drawn by assuming that the branch ratio between the spreading width and
the $\pi$N decay is roughly one.  However, our calculation in this work
show that the branching
ratio has a clear momentum dependence which should be properly taken into 
account.  In this regard,
our results, especially the results at saturation density, could provide
important constraints. 

In summary, we have calculated the delta spreading width in the framework
of a relativistic nucleon-meson model.  
Our concern is to present the branching ratio of the 
spreading width to $\pi$N decay 
so that it can be used for the analysis  of  measurements 
of the nucleon axial form factor~\cite{BNL} or for the 
atmospheric neutrino experiments~\cite{kamiokande}.  
$\pi$N decay is substantially reduced by Pauli blocking and mean fields. 
Also, mean fields play an important role in the spreading width.
We have found that the spreading width is large even though its
relative strength to the $\pi$N decay depends on the nuclear
density and the relative motion of the delta with respect to the nuclear
medium.

\acknowledgments
SHL was supported in part by the by the Basic Science Research
Institute program of the Korean Ministry of Education through grant no.
BSRI-96-2425, by KOSEF through the CTP at Seoul National University
and by Yonsei University Research Grant.  
HK  was supported by KOSEF.  We thank C. J. Horowitz for useful discussions
in the early stage of this work.

\begin{figure}
\caption{Diagrams for $\Delta$ self-energy.  Figure (a) is the self-energy
diagram representing the $\pi$N decay and figure (b) is the lowest diagram for
the spreading width.}
\label{self}
\end{figure}
\begin{figure}
\caption{Pionic decay of $\Delta$ with respect to the delta momentum.  
(a) is for the $\pi$N decay without mean fields and (b) 
includes the mean fields. 
The solid lines show the results for the half saturation density, 
the dashed lines
for the saturation density and the dot-dashed lines for twice the 
saturation density.}
\label{wid}
\end{figure}
\begin{figure}
\caption{Spreading width of $\Delta$. 
(a) shows the result without mean fields and (b) includes the
mean fields.  
The solid lines show the results for the half saturation density, 
the dashed lines
for the saturation density and the dot-dashed lines for twice the  
saturation density.}
\label{swid}
\end{figure}
\eject

\setlength{\textwidth}{6.1in}   
\setlength{\textheight}{9.in}  
\begin{figure}
\centerline{%
\vbox to 2.4in{\vss
   \hbox to 3.3in{\includegraphics{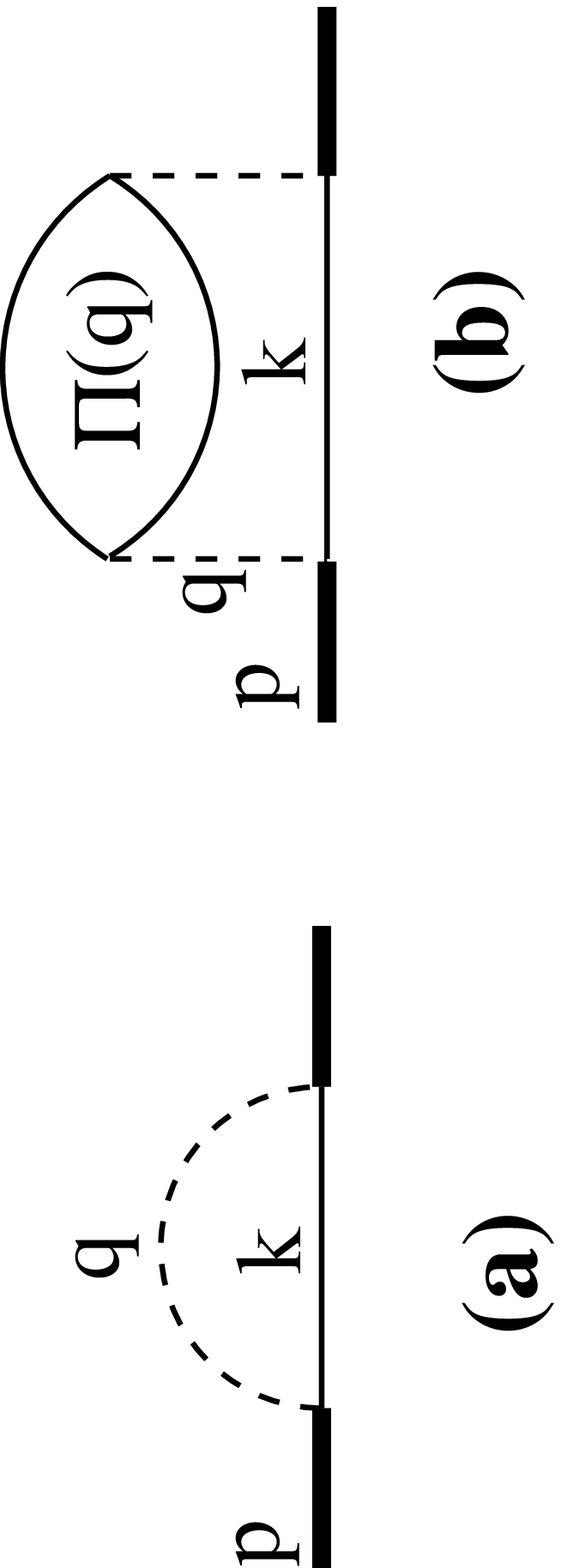}\hss}}
}
\bigskip
\vspace{400pt}
Figure 1
\end{figure}
\eject
\begin{figure}
\centerline{%
\vbox to 2.4in{\vss
   \hbox to 3.3in{\includegraphics{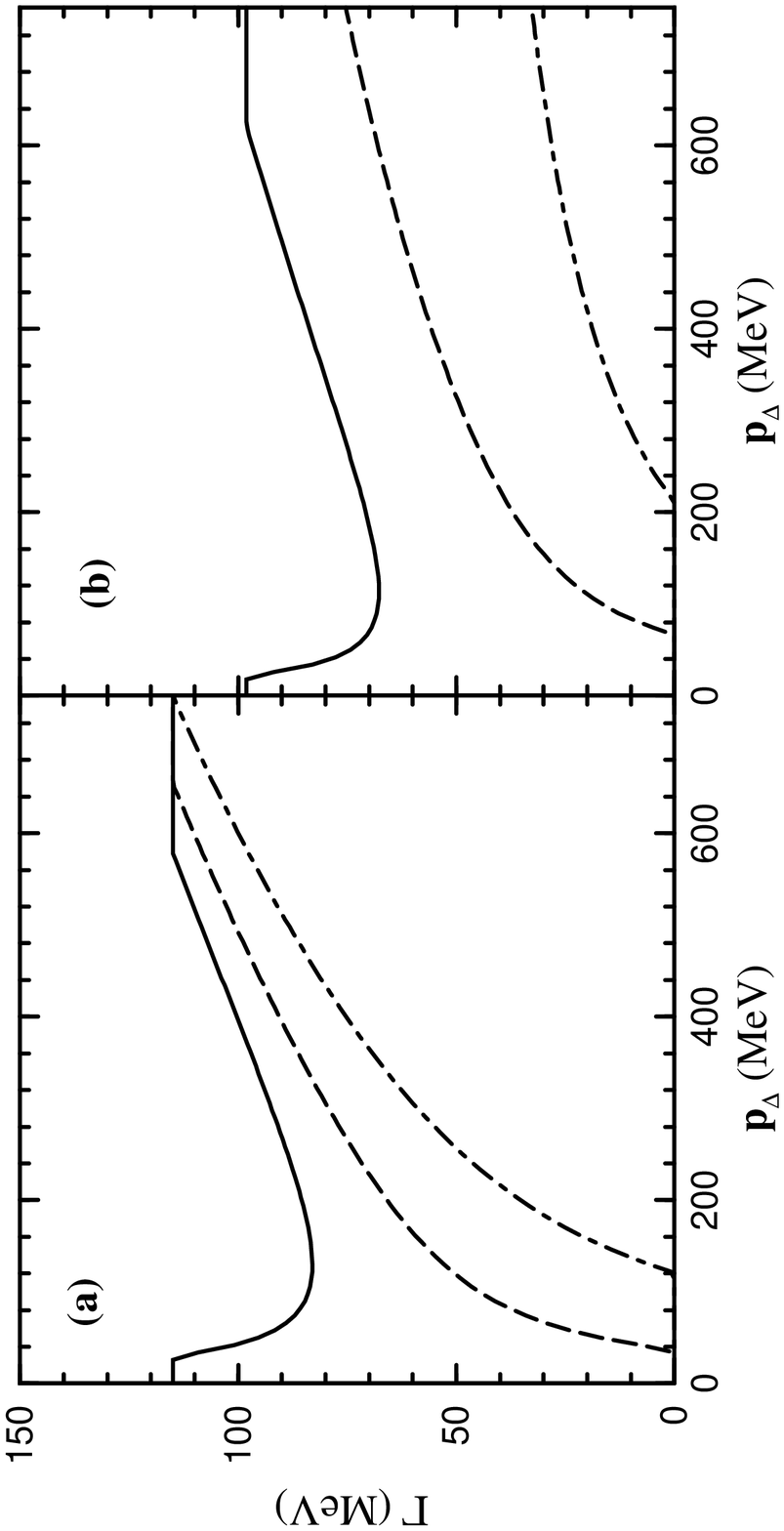}\hss}}
}
\bigskip
\vspace{400pt}
Figure 2 
\end{figure}
\eject
\begin{figure}
\centerline{%
\vbox to 2.4in{\vss
   \hbox to 3.3in{\includegraphics{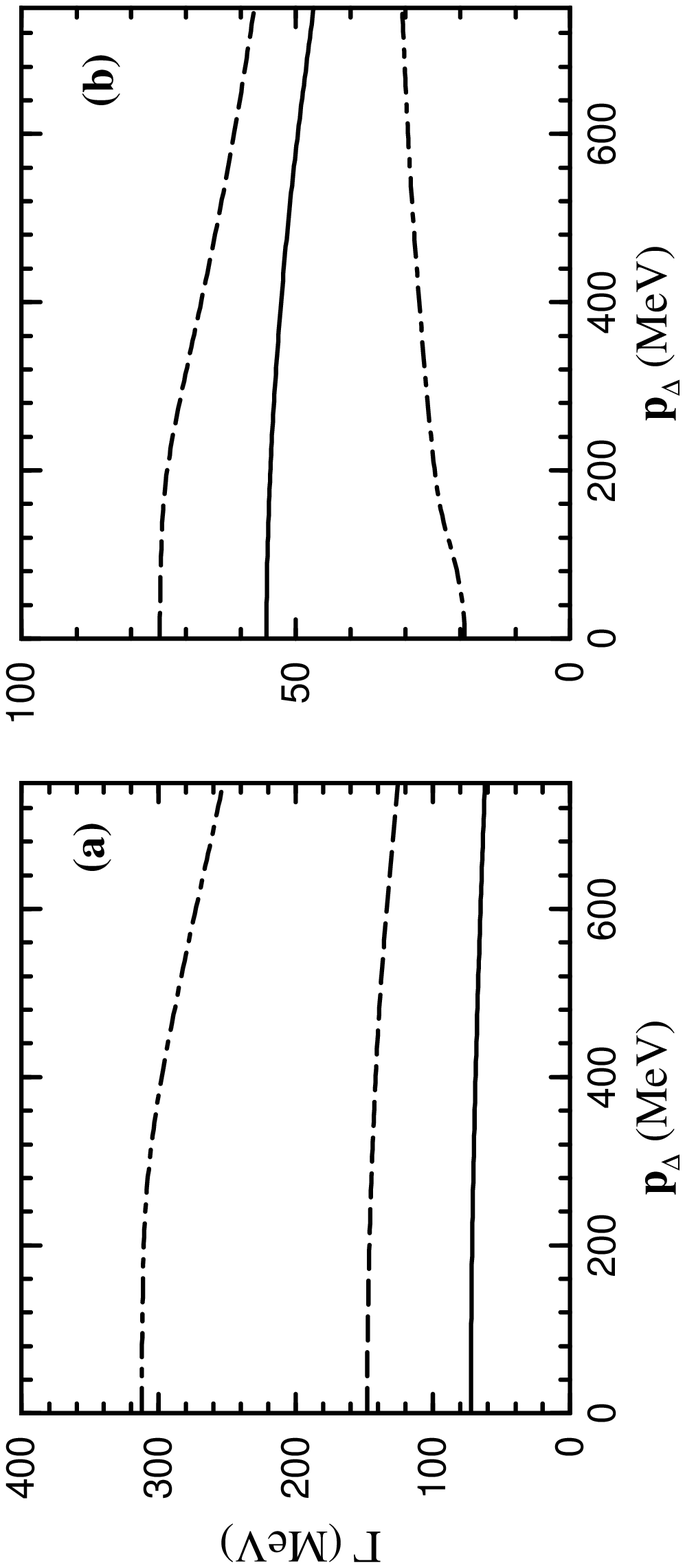}\hss}}
}
\bigskip
\vspace{400pt}
Figure 3 
\end{figure}
\end{document}